# Demonstration of high output power (3.75 W/mm) with 31.8% PAE in Ga-Polar AlGaN/GaN HEMTs on sapphire at 94 GHz

Ruixin Bai, Swarnav Mukhopadhyay, Michael Elliott, Ryan Gilbert, Surjava Sanyal, Darryl Shima, Jiahao Chen, Rafael A. Choudhury, Kyudong Kim, Yu-Chun Wang, Ganesh Balakrishnan, Ahmad E. Islam, Andrew J. Green, Shubhra S. Pasayat and Chirag Gupta

*Abstract*— **In this work, a recessed-gate Ga-polar AlGaN/GaN high-electron-mobility transistor (HEMT) on sapphire incorporating an AlGaN back-barrier for W-band operation is demonstrated. The device achieves a subthreshold slope of ~130 mV/dec, a peak transconductance of 0.75 S/mm, and an fmax exceeding 240 GHz. Continuous-wave load-pull measurements at 94 GHz show an output power of 3.75 W/mm with a power-added efficiency of 31.8%. This performance significantly exceeds the best previously reported Ga-polar devices on sapphire and approaches that of state-of-the-art Ga-polar devices on SiC, which provides a promising pathway toward high-performance W-band devices.**

## I. INTRODUCTION

Millimeter-wave operation at the W-band (75–110 GHz) has attracted significant interest in recent years due to increasing spectrum congestion at lower frequencies and the growing demand for higher bandwidth in advanced radar and communication systems. GaN high-electron-mobility transistors (HEMTs) are attractive candidates for W-band applications due to their high electron mobility and high breakdown field, enabling high-power density and high-efficiency operation at millimeter-wave frequencies.[1] Meanwhile, GaN HEMTs on sapphire substrates offer a cost-effective and scalable platform compatible with 8-inch and larger wafers, making them attractive for large-scale manufacturing.

In 2024, Collins et al. demonstrated recessed-gate N-polar GaN HEMTs on sapphire substrates with an output power ($P_{OUT}$) of 3.7 W/mm and a power-added efficiency (PAE) of 53% at 94 GHz. [2] However, N-polar growth and device fabrication typically involve increased complexity compared with conventional Ga-polar HEMT technologies. [3] In contrast, only limited studies have investigated Ga-polar GaN HEMTs on sapphire for W-band operation and the reported performance is generally much lower than that of N-polar counterparts, with typical $P_{OUT}$ of around 1.5 W/mm and PAE of about 15%. [4] Recently, an AlGaN/GaN HEMT on a sapphire substrate with a recessed-gate design has been demonstrated. Benefiting from a passivation-free design and improved dispersion control enabled by the recessed-gate structure, the device achieved a significantly improved $P_{OUT}$ of 2.8 W/mm with a PAE of 26.8%. [5] However, compared with N-polar devices, the absence of a back barrier leads to pronounced short-channel effects, which still limit the achievable $P_{OUT}$ and PAE. Nevertheless, back barrier can also be incorporated into Ga-polar HEMTs. An early study by Palacios et al. introduced an InGaN back barrier to enhance 2DEG confinement through polarization engineering. [6] AlGaN back barriers were subsequently explored as a practical alternative, with Lee et al. demonstrating their effectiveness in suppressing short-channel effects, thereby improving the RF performance of InAlN/GaN HEMTs. [7] Subsequent studies further demonstrated that AlGaN back barriers can also improve dispersion control and breakdown characteristics, enabling 3.5 W/mm output power and 66.5% PAE at 40 GHz. [8] More recently, Park et al. demonstrated an $f_{max}$ of 411 GHz in AlGaN/GaN HEMTs incorporating an AlGaN back barrier. [9] Despite these advances, the impact of AlGaN back-barrier on the large-signal performance of W-band remains relatively unexplored. Moreover, its compatibility with recessed-gate structures employing thick AlGaN top barriers, which have recently demonstrated promising potential for W-band operation, has not yet been systematically investigated.

In this work, an AlGaN back barrier is introduced beneath the UID GaN channel in the recessed-gate structure, and its impact on the W-band performance is systematically investigated. The incorporation of the back barrier effectively suppresses short-channel effects, enabling the device with 90 nm gate length to achieve a subthreshold slope below 150 mV/dec, a peak transconductance exceeding 0.75 S/mm, and a

This work was supported by DEPSCOR AFOSR under Grant FA9550-24-1-0159 and AFRL Regional Network – Midwest. (*Corresponding author: Ruixin Bai.)*

Ruixin Bai, Swarnav Mukhopadhyay, Surjava Sanyal, Jiahao Chen, Rafael A. Choudhury, Kyudong Kim, Yu-Chun Wang, Shubhra S. Pasayat and Chirag Gupta are with Department of Electrical and Computer Engineering, University of Wisconsin-Madison, Madison, WI 53706 USA. (e-mail: rbai33@wisc.edu).

Michael Elliott and Ryan Gilbert are with KBR Inc, Wright-Patterson AFB, Dayton, OH 45433 USA.

Darryl Shima and Ganesh Balakrishnan are with Center for High Technology Materials, University of New Mexico, Albuquerque, NM 87106 USA.

Ahmad E. Islam and Andrew J. Green are with Air Force Research Laboratory Sensors Directorate, Wright-Patterson AFB, OH 45433 USA

maximum oscillation frequency ($f_{max}$) above 240 GHz. As a result, the device delivers a $P_{OUT}$ of 3.75 W/mm with a PAE of 31.8% at 94 GHz. These results exceed the best previously reported Ga-polar devices on sapphire by more than 30% and approach the performance of state-of-the-art Ga-polar devices fabricated on SiC substrates. In addition, under comparable drain bias conditions, the achieved $P_{OUT}$ approaches the performance reported for N-polar devices. Considering the low-cost sapphire substrate and the reduced fabrication complexity, this approach provides a promising pathway toward high-performance W-band power devices.

## II. Device Fabrication

The epitaxial structure used in this work was grown by metal-organic chemical vapor deposition (MOCVD) on a 2-inch Fe-doped GaN-on-sapphire substrate. The layer design mainly follows that reported in Ref.[10], with several modifications. In particular, a 15 nm $Al_{0.12}Ga_{0.88}N$ back barrier was introduced above the GaN buffer layer, and the GaN channel thickness was reduced to 120 nm. As shown in **Fig. 1(a)**, the final structure consists of a 15 nm $Al_{0.12}Ga_{0.88}N$ back barrier, a 120 nm unintentionally doped (UID) GaN channel, a 0.7 nm AlN interlayer, and a 31 nm $Al_{0.35}Ga_{0.65}N$ barrier.

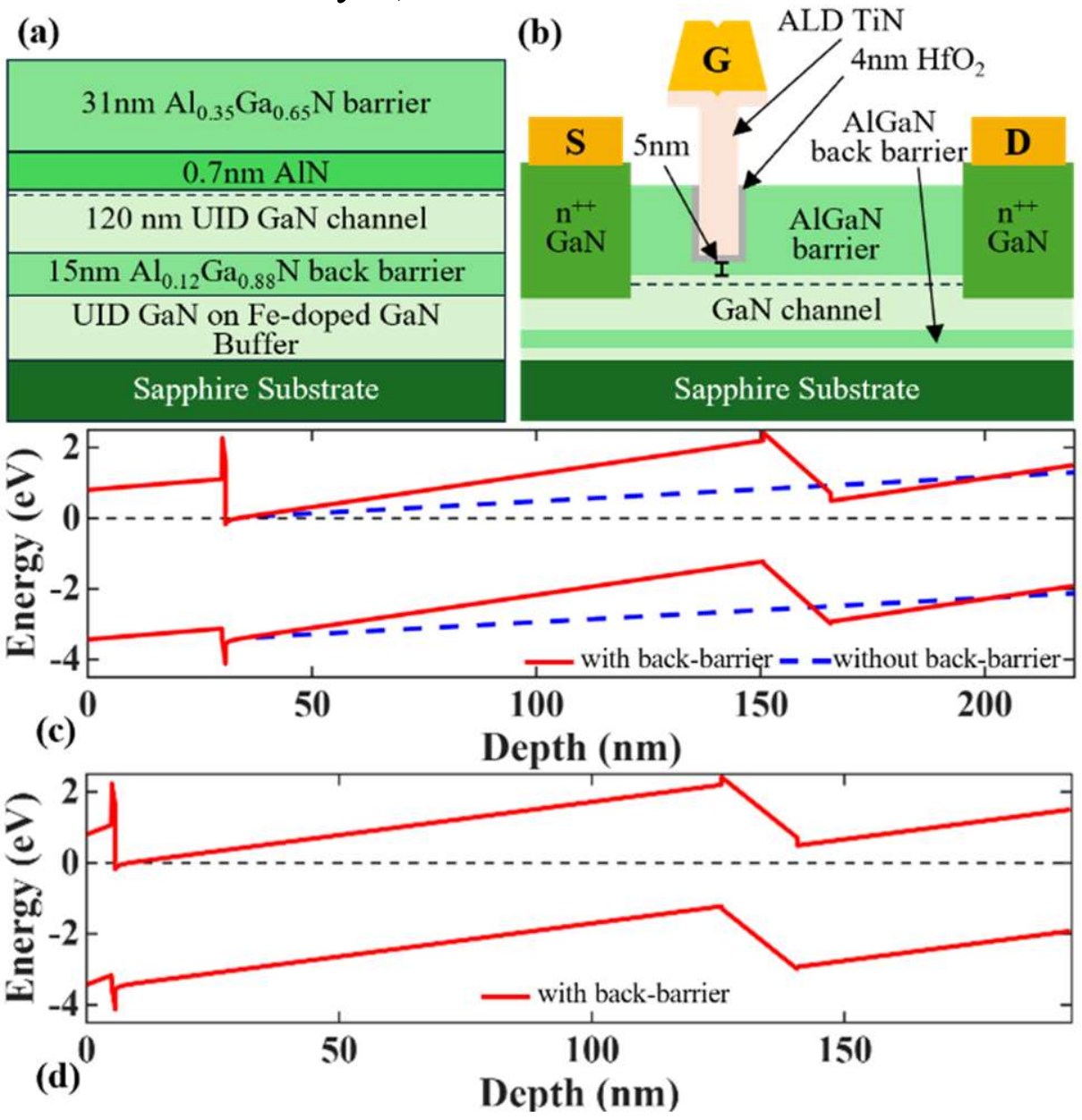

Fig. 1. (a) Epitaxial layer structure and (b) cross-section schematic view of the recessed-gate AlGaN/GaN HEMT with back barrier design. (c) Simulated band diagrams for devices with and without back-barrier in access region. (d) Simulated band diagrams under the gate.

As shown in **Fig. 1(c)**, comparison of the band diagrams reveals enhanced electron confinement in the 2DEG region with the incorporation of the AlGaN back-barrier, which suppresses electron spillover into the buffer layer and thus mitigates short-channel effects. Device fabrication began with the selective area regrowth of a 90-nm n+ GaN layer in the ohmic regions using MOCVD, with an electron concentration of ~2 × $10^{20}$ $cm^{-3}$. [11] Following regrowth, mesa isolation was carried out using a $BCl_3/Cl_2$-based reactive ion etching (RIE) process. A 200 nm $SiO_2$ layer was then deposited by plasma-enhanced chemical vapor deposition (PECVD) and served as a hard mask. The gate foot region was subsequently defined by electron-beam lithography (EBL) using ARP6200 resist, followed by etching of the exposed $SiO_2$ in an ICP system with $CHF_3/CF_4/O_2$ plasma. The AlGaN barrier in the gate region was then recessed using a low-power $BCl_3/Cl_2$-based RIE process, leaving approximately 5 nm of AlGaN under the gate to increase the gate aspect ratio, as shown in **Fig. 1(b)**. Prior to gate metal deposition, a 4 nm $HfO_2$ layer was deposited by atomic layer deposition (ALD) at 250 °C to serve as the gate dielectric. Subsequently, a 50 nm TiN gate metal was deposited by ALD at 275 °C with $N_2$ plasma assistance. The T-gate head was formed in a second EBL step and metallized with Cr/Au using electron-beam evaporation. After gate formation, the $SiO_2$ hard mask was completely removed by buffered oxide etchant (BOE) wet etching. Finally, the ohmic contacts and thick probing pads were deposited by electron-beam evaporation. The fabricated devices feature a T-gate with a foot length ($L_g$) of 90 nm, a gate top length of 600 nm, a source-to-gate spacing ($L_{sg}$) of 75 nm, source-to-drain spacing ($L_{sd}$) of 400 nm and a total gate width ($W_g$) of 2 × 37.5 µm.

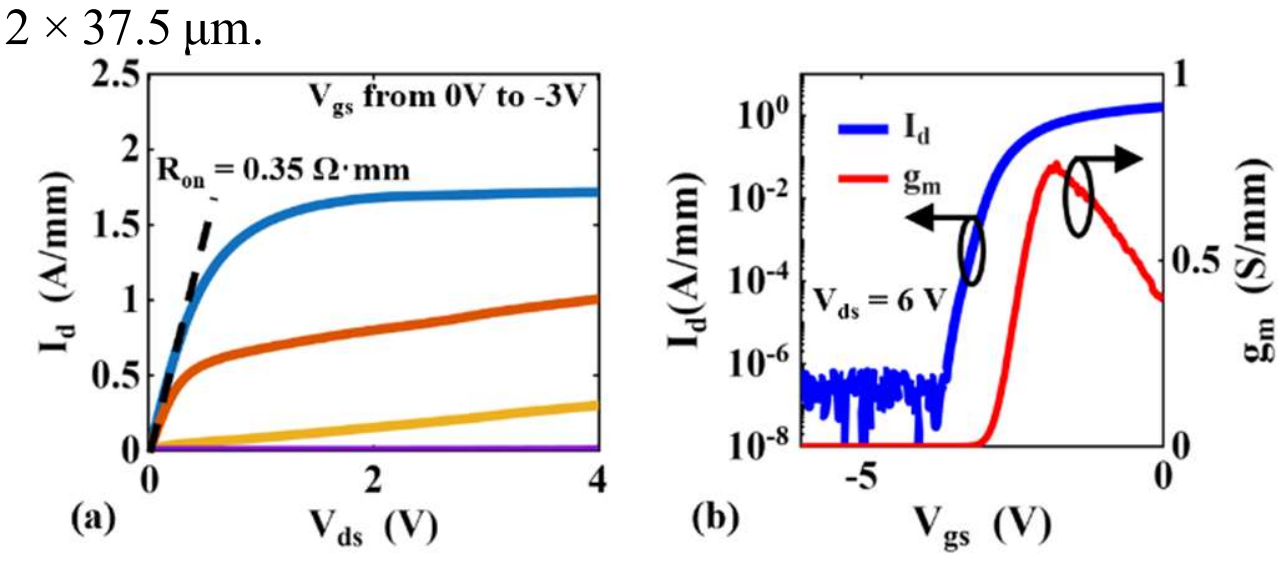

Fig. 2. (a) DC output characteristics and (b) transfer characteristics measured at $V_{ds}$ = 6 V of the device with back-barrier.

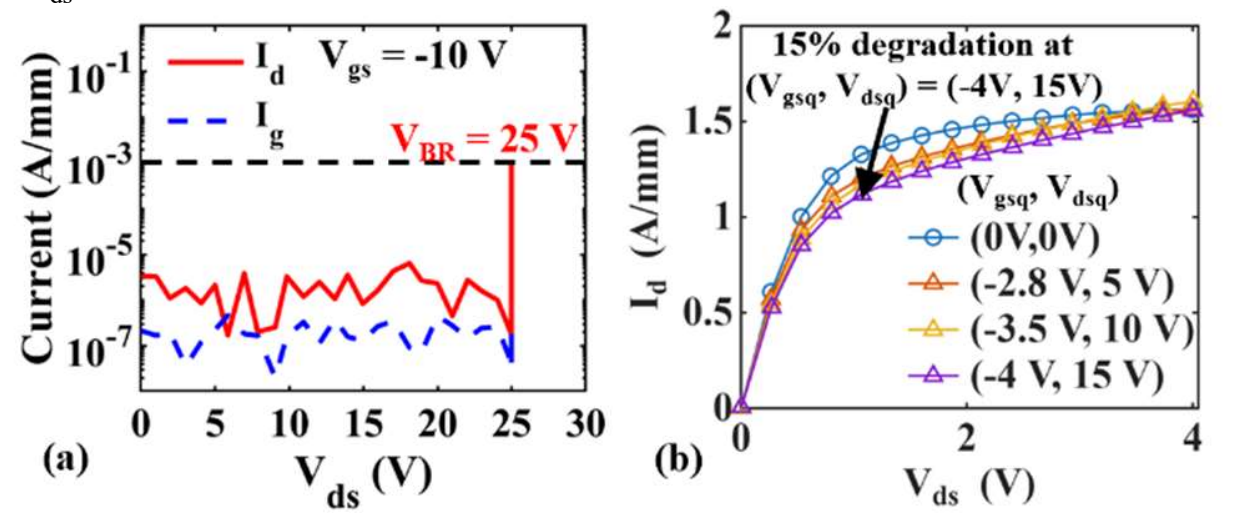

Fig. 3. (a) Three-terminal off-state breakdown characteristics of the device. (b) Pulsed I–V characteristics of the device with back barrier measured at ($V_{gsq}$, $V_{dsq}$) of (-2.8 V, 5 V), (-3.5 V, 10 V) and (-4 V, 15 V), with $V_{gsq}$ adjusted to maintain an $I_{dsq}$ of 250 mA/mm.

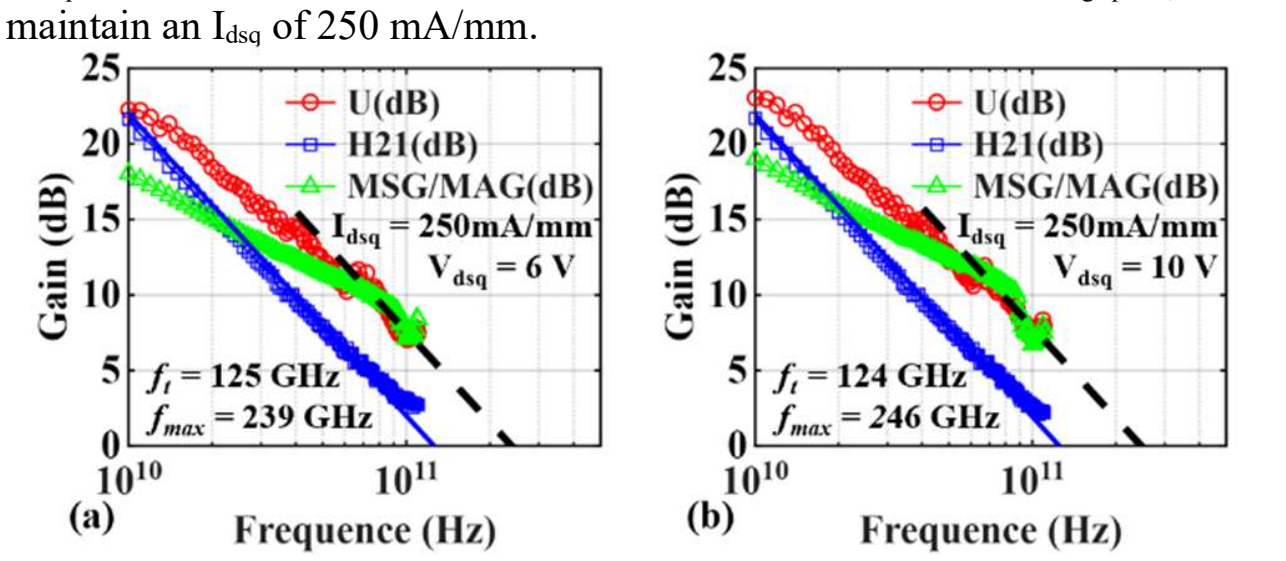

Fig. 4. H21, U and MSG/MAG vs frequency measured at (a) $I_{dsq}$ = 250 mA/mm and $V_{dsq}$ = 6 V and (b) $I_{dsq}$ = 250 mA/mm and $V_{dsq}$ = 10 V.

## III. Results and Discussion

**Fig. 2(a)** shows the DC output characteristics of the fabricated device. A maximum drain current density of 1.7 A/mm is achieved, together with a low on-resistance ($R_{on}$) of 0.35 Ω·mm. The high current density and low $R_{on}$ are enabled by the regrown ohmic contacts, which exhibit a contact resistance around 0.1 Ω·mm extracted from TLM measurements. [11] The $R_{sh}$ of the access region is approximately 260.4 Ω/□, while the $R_{sh}$ of the region under the gate is around 445 Ω/□. The transfer characteristics measured at $V_{ds}$ = 6 V are shown in **Fig. 2(b)**. A peak transconductance ($g_m$) of 0.75 S/mm and a high on/off current ratio of $10^7$ are achieved. Benefiting from the incorporation of an AlGaN back barrier to suppress short-channel effects, the device exhibits a subthreshold slope of ~130 mV/dec, which is significantly lower than that of previously reported W-band recessed-gate devices without a back barrier. [5] The relatively negative threshold voltage of -3V, despite the gate recess, is attributed to the combined effects of the relatively high-Al-composition AlGaN top barrier, the lower-work-function TiN gate, and the MISHEMT gate structure, which help preserve 1.2e13cm$^{-2}$ charge density at $V_{gs}$ = 0V under the gate region. In addition, the $V_{th}$ values are relatively uniform across the devices, with only a small variation observed for different $L_g$. Further improvement in $V_{th}$ uniformity could be achieved through more precise gate recess control, such as by employing an AlN etch-stop layer or using atomic layer etching.

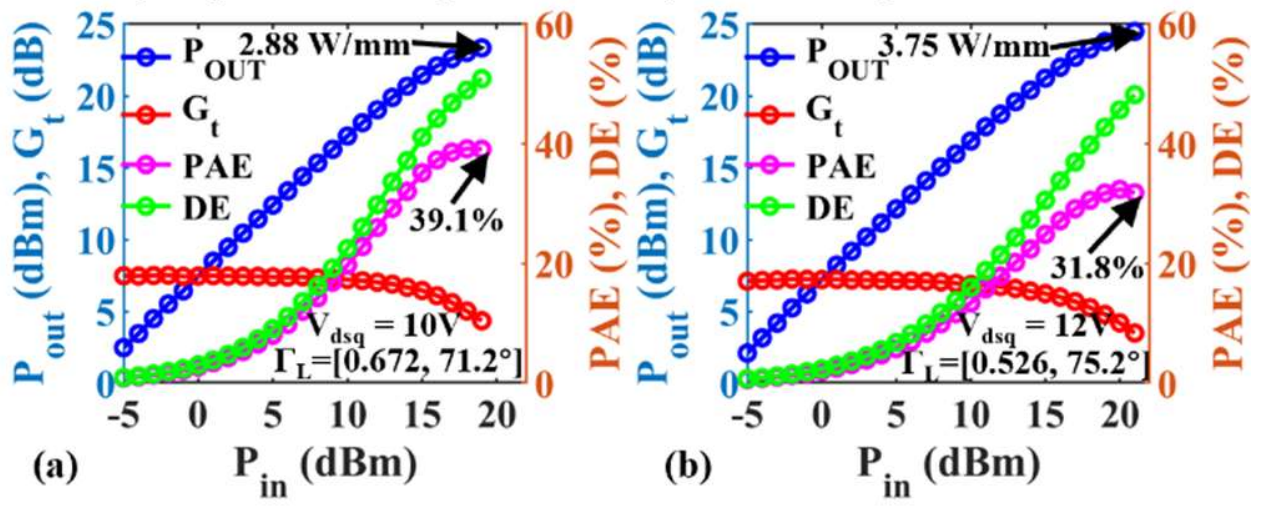

Fig. 5. 94GHz load pull power sweep in CW mode with $I_{dsq}$ = 250mA/mm at (a) $V_{dsq}$ = 10 V and (b) $V_{dsq}$ = 12 V.

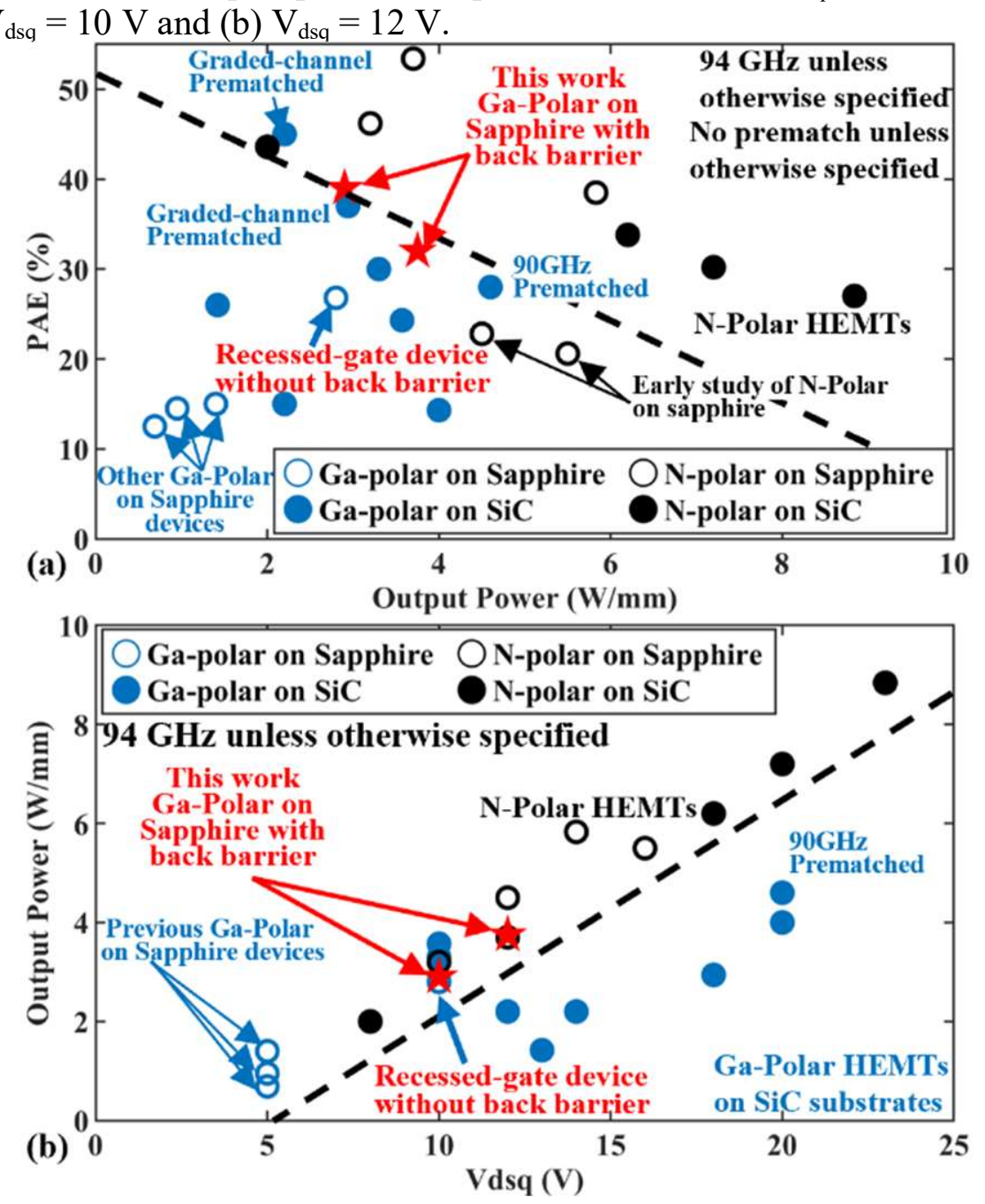

Fig. 6. (a) Benchmark of PAE versus associated $P_{OUT}$ at 83-95 GHz for single-channel GaN HEMTs. (b) $P_{OUT}$ versus $V_{dsq}$ benchmark of single-channel GaN HEMTs measured at 83-95 GHz.

**Fig. 3(a)** shows the three-terminal off-state breakdown characteristics, yielding a breakdown voltage of 25 V at $V_{gs}$ = −10 V, corresponding to an average breakdown field of 1.06 MV/cm. Future optimization of the AlGaN back-barrier, gate-recess process, and gate-dielectric deposition could potentially increase the breakdown voltage, providing greater voltage margin for load-pull operation. Pulsed I–V measurements were performed with a pulse width of 50 µs and a duty cycle of 1% under semi-on bias conditions corresponding to an $I_{dsq}$ of 250 mA/mm, with $V_{dsq}$ varied to represent Class-AB operation. It should be noted that the pulse width (50 µs) was limited by the tool's capability. As shown in **Fig. 3(b)**, when compared at $V_{ds}$ = 1 V, the device exhibits a current collapse of 12% at ($V_{gsq}$, $V_{dsq}$) of (-2.8 V, 5 V) and 15% at ($V_{gsq}$, $V_{dsq}$) of (-4 V, 15 V), which remains within a reasonable range compared with other Ga-polar W-band devices. [12], [13] This behavior is mainly attributed to the recessed-gate structure, which introduces a threshold voltage difference between the gate region and the gate access region. As a result, the virtual gate effect induced by surface traps in the access region is partially compensated, leading to reduced current dispersion. [5], [14] In addition, the recessed-gate structure enables our passivation-free device design, which reduces parasitic capacitance and is beneficial for high-frequency operation.

TABLE I
EXTRACTED SMALL-SIGNAL PARAMETERS

| | This work back barrier | | Ref.[5] no back barrier | |
|---|---|---|---|---|
| $V_{dsq}$ | 6 V | 10 V | 6 V | 10 V |
| $R_s$ (Ω) | 1.60 | 1.60 | 1.83 | 1.83 |
| $R_d$ (Ω) | 2.16 | 2.16 | 2.17 | 2.17 |
| $g_m$ (mS) | 56.3 | 54.4 | 51.4 | 41.7 |
| $C_{gs}$ (fF) | 56.4 | 57.6 | 51.1 | 46.8 |
| $C_{gd}$ (fF) | 10.1 | 8.3 | 9.6 | 9.2 |
| $R_{ds}$ (Ω) | 334 | 294 | 250 | 290 |
| $R_g$ (Ω) | 2.6 | 2.6 | 3.1 | 3.1 |
| $F_{t_measured}$ (GHz) | 125 | 124 | 127 | 112 |
| $F_{max_measured}$ (GHz) | 239 | 246 | 203 | 191 |
| $F_{t_model}$ (GHz) | 129 | 126 | 129 | 114 |
| $F_{max_model}$ (GHz) | 231 | 241 | 202 | 192 |

S-parameters were measured from 10GHz to 110 GHz in a single sweep using mTRL calibration. As shown in **Fig. 4**, the RF performance was evaluated under two quiescent drain bias conditions while the quiescent drain current was fixed at 250 mA/mm, corresponding to the device operating point. At $V_{dsq}$ = 6 V, the device achieves a $f_t$ of 125 GHz and a $f_{max}$ of 239 GHz. When $V_{dsq}$ is increased to 10 V, $f_t$ remains nearly unchanged at 124 GHz, while $f_{max}$ further improves to 246 GHz, which is 25% higher than the previous device without back barrier. To further understand the origin of this improvement, key small-signal parameters were extracted and compared with those of the device without the back barrier. At $V_{dsq}$ = 6 V, the present device exhibits slightly higher $C_{gs}$, $C_{gd}$, and $g_m$, primarily due to the slightly thinner barrier. The $R_{ds}$ is also approximately 35% higher, indicating reduced output conductance and suppressed short-channel effects. More importantly, at $V_{dsq}$ = 10 V, the device without the back barrier exhibits an approximately 20% reduction in $g_m$ due to stronger short-channel effects, whereas $g_m$ degrades by less than 5% in the present device with back barrier, indicating improved gate control enabled by the AlGaN back barrier. Meanwhile, $C_{gd}$ decreases by approximately 20% with increasing $V_{dsq}$ due to the expansion of the drain-side depletion region. The preserved gm and reduced $C_{gd}$ together lead to the enhanced $f_{max}$ of 246 GHz at $V_{dsq}$ = 10 V, leading to the higher small-signal gain at 94 GHz. With device scaling to $L_g$ < 50 nm and $L_{sd}$ < 300 nm in the future, together with optimization of the back-barrier design, both $f_t$ and $f_{max}$ are expected to improve, bringing the device closer to the requirements for W-band operation.

Continuous-wave (CW) load-pull measurements were performed at 94 GHz using a scalar passive load-pull system, with TRL calibration carried out prior to the measurements. [15] As shown in **Fig. 5**, at $V_{dsq}$ = 10 V and $I_{dsq}$ = 250mA/mm, the device achieves a $P_{OUT}$ of 2.88 W/mm with a PAE of 39.1%. When $V_{dsq}$ is increased to 12 V, the $P_{OUT}$ further improves to 3.75 W/mm with a PAE of 31.8%. To the best of our knowledge, both the achieved $P_{OUT}$ and PAE at 94 GHz represent the highest reported values for Ga-polar GaN channel HEMTs on sapphire substrates. [4], [5]

**Fig. 6** provides a comparison of this work with the previously reported recessed-gate device without back barrier [5] and other Ga-polar and N-polar GaN HEMTs included in the benchmark in Ref.[5]. As illustrated in **Fig. 6(a)**, when compared with the device without back barrier, at the same $V_{dsq}$= 10 V and comparable $P_{OUT}$, the present back-barrier device achieves nearly 50% higher PAE, owing to the suppressed short-channel effects that preserve $f_{max}$ and enhance the gain at 94 GHz. In addition, the AlGaN back barrier improves the high-field robustness and breakdown characteristics of the HEMT, enabling operation at a higher $V_{dsq}$ and consequently yielding approximately 30% higher maximum $P_{OUT}$, bringing the performance closer to that of the best Ga-polar devices reported on SiC substrates. Furthermore, as shown in **Fig. 6(b)**, the device in this work delivers higher $P_{OUT}$ than most reported Ga-polar devices under comparable bias conditions. In particular, the performance obtained at $V_{dsq}$ = 12 V, together with the observed $P_{OUT}$ scaling behavior with increasing drain bias, reaches a level comparable to that of reported N-polar devices. Taken together, considering the use of a low-cost sapphire substrate and the relatively simple fabrication process, these results highlight the strong potential of recessed-gate Ga-polar AlGaN/GaN HEMTs on sapphire for next-generation millimeter-wave power applications.

## IV. CONCLUSION

In summary, a recessed-gate Ga-polar AlGaN/GaN HEMT on sapphire substrate incorporating an AlGaN back barrier was demonstrated for W-band RF applications in this work. The back barrier effectively suppresses short-channel effects and improves electrostatic control, leading to excellent DC and RF performance. The device exhibits a $f_t$ of 125 GHz and an $f_{max}$ up to 246 GHz with a small signal gain of more than 8dB at 94GHz. Continuous-wave load-pull measurements at 94 GHz achieve a $P_{OUT}$ of 3.75 W/mm with an associate PAE of 31.8% at $V_{dsq}$ = 12 V. Benchmarking results show that the achieved $P_{OUT}$ and PAE represent the highest reported values for Ga-polar GaN channel HEMTs on sapphire substrate, highlighting the strong potential of this technology for future W-band power applications.